\documentclass[fleqn]{revtex4}

\usepackage{amsmath,graphicx}

\begin{document}

\title{Calculation of the nonrelativistic Bethe logarithm in the velocity gauge}

\author{V.I.~Korobov}
\affiliation{Joint Institute for Nuclear Research, 141980, Dubna, Russia}

\begin{abstract}
We consider a general procedure to evaluate the Bethe logarithm for a general few-body atomic or molecular system. As benchmarks we use calculation for the ground states of a helium atom and H$_2^+$ molecular ion. The obtained values are: $\beta_{\rm He}=4.37016022306(2)$ for the helium atom and $\beta_{{\rm H}_2^+} = 3.012230335(1)$ for the H$_2^+$. Both results substantially improve the best known values for these quantities.
\end{abstract}

\maketitle

\section{Introduction}

It is known that in the leading order radiative contribution to the binding energy of atomic or molecular system the most complicate for numerical evaluation quantity is the Bethe logarithm \cite{BS}. One of the first most accurate results for the helium ground state was obtained by C.~Schwartz \cite{Schwartz} in 1961 and remained the best one over 30 years! In 1999, Goldman and Drake \cite{Dra99} suggested a new way to evaluate the Bethe logarithm, $\beta(n,L)$, for a helium atom, which is based on another presentation of $\beta(n,L)$ in terms of the {\em acceleration gauge} dipole operators and a full diagonalization of the Hamiltonian. This method works well for two- and three-electron helium(lithium)-like atoms \cite{DY} but attempts to apply it to other systems like H$_2^+$ molecular ion were not very successful. The other disadvantage of the method is a necessity to add some extra terms into a basis set, which looks like $1/r$ times regular solution.

The major aim of a present work is to elaborate a universal method, which uses definition of the Bethe logarithm in terms of dipole operators in the {\em velocity gauge} and still as efficient as the Goldman-Drake method in case of hydrogen and helium like atoms. In fact, we have tried to carefully reanalyze the ideas of \cite{Schwartz} and to present them in a more explicit and general form. The numerical results are confined to three-body systems while the theoretical expressions are valid for a general few-body case.

The paper is organized as follows. In Section II we consider a derivation of the leading order radiative corrections that allow us to define the Bethe logarithm for a general few-body Coulomb system as an integral over photon energy $k$. Next the asymptotic behaviour of the core integrand, $J(k)=\left\langle       \mathbf{J}\left(E_0\!-\!H\!-\!k\right)^{-1}\mathbf{J} \right\rangle$, and its first order perturbation wave function $\psi_1(k)$, are derived, the leading order terms are obtained as expectation values of some operators. A variational property of the integral over $k$ is discussed. It allows us to work out an efficient numerical scheme to calculate the Bethe logarithm using optimal parameters, which are variationally chosen.

In Section III the numerical method, which is the main goal of our studies, is describe in details. And finally, the results of the Bethe logarithm calculation for the hydrogen, helium and $\mathrm{H}_2^+$ molecular ion ground states are presented and compared with other available results.

We are using atomic units ($\hbar=e=m_e=1$) throughout if something else is not explicitly stated.

\section{Theoretical overview}

\subsection{Radiative correction. Low energy contribution}

Let us first consider a case of a hydrogen-like atom.

\begin{figure}
\includegraphics[width=0.24\textwidth]{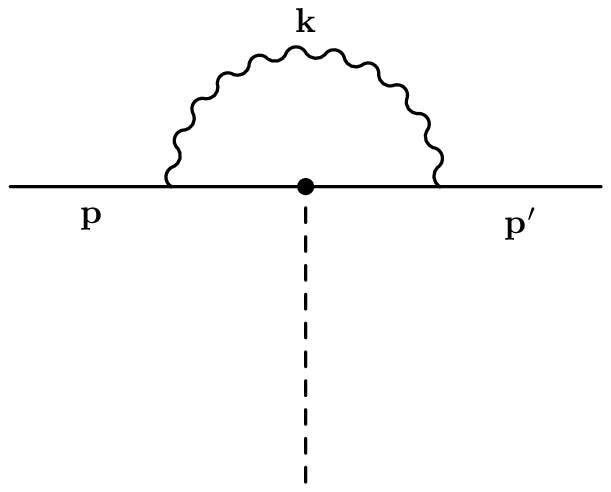}(a)\hspace{8mm}
\includegraphics[width=0.24\textwidth]{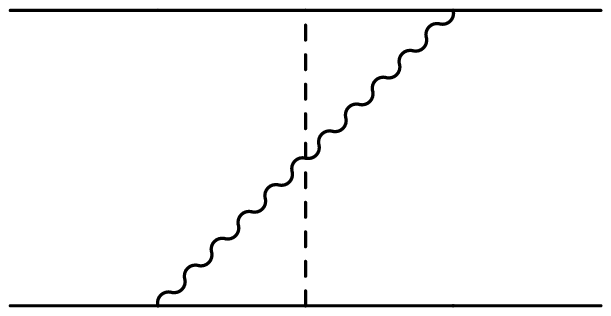}\hspace{2mm}(b)
\caption{NRQED diagrams for the self-energy and retarded transverse photon at ultra-low energies}\label{NRQED1}
\end{figure}

The $\alpha(Z\alpha)^2E_{nr}$ order low-energy contribution, which results from the NRQED diagram (see Fig.~\ref{NRQED1}\,(a)), may be written
\begin{equation}\label{SE_1}
\begin{array}{@{}l}\displaystyle
E_{L} =
   \frac{\alpha^3}{4\pi^2m^2}\int_{|\mathbf{k}|<\Lambda}\frac{d^3k}{k}
   \left(\delta^{ij}-\frac{k^ik^j}{k^2}\right)
   \left\langle
      \psi_0\left|p^i\left(\frac{1}{E_0-H-k}\right)p^j\right|\psi_0
   \right\rangle
 - \delta m\left\langle\psi_0|\psi_0\right\rangle.
\end{array}
\end{equation}
where the last term is the "mass renormalization" contribution. Averaging the integrand over angular variables one gets
\begin{equation}
E_{L} =
   \frac{2\alpha^3}{3\pi m^2}\int_0^\Lambda\,k\,dk
   \left\langle
      \mathbf{p}\left(\frac{1}{E_0-H-k}\right)\mathbf{p}
   \right\rangle - \delta m\left\langle\psi_0|\psi_0\right\rangle.
\end{equation}

The integrand may be rewritten using the following operator identity
\[
(E_0\!-\!H\!-\!k)^{-1} =
   -1/k-\frac{1}{k^2}(E_0\!-\!H)+\frac{1}{k^2}\frac{(E_0-H)^2}{E_0\!-\!H\!-\!k}
\]
that results in
\begin{equation}\label{eq3}
E_{L} =
   \frac{2\alpha^3}{3\pi m^2}
   \left[
      {-\left\langle
         \mathbf{p}^2
      \right\rangle \Lambda}
      +\left\langle
         \mathbf{p}\left[H,\mathbf{p}\right]
      \right\rangle \ln{\Lambda}
      +\int\,\frac{dk}{k}\,
      \left\langle
         \mathbf{p}\frac{(E_0\!-\!H)^2}{E_0\!-\!H\!-\!k}\mathbf{p}
      \right\rangle
   \right] - \delta m\left\langle\psi_0|\psi_0\right\rangle.
\end{equation}
As was shown by Bethe in 1947 \cite{Bet47}, the linearly divergent term should be associated with the "mass renormalization"\/ of an electron and should be subtracted with the last term in expression (\ref{eq3}). Thus, the remaining part may be splitted onto a finite nonlogarithmic contribution
\begin{subequations}\label{SE_fin}
\begin{equation}
\begin{array}{@{}l}\displaystyle
E_{L}^{(0)} =
   \frac{2\alpha^3}{3\pi m^2}\int_0^{E_h}\,k\,dk
   \left\langle
      \mathbf{p}\left(\frac{1}{E_0-H-k}+\frac{1}{k}\right)\mathbf{p}
   \right\rangle
   +\frac{2\alpha^3}{3\pi m^2}\int_{E_h}^\infty\,\frac{dk}{k},
   \left\langle
      \mathbf{p}\frac{(E_0\!-\!H)^2}{E_0\!-\!H\!-\!k}\mathbf{p}
   \right\rangle
\end{array}
\end{equation}
and the divergent part
\begin{equation}
E_{L}^{(1)} =
   \frac{2\alpha^3}{3\pi m^2}\left( \int_{E_h}^{\Lambda}\,\frac{dk}{k} \right)
   \Bigl\langle
      \mathbf{p}\left[H,\mathbf{p}\right]
   \Bigr\rangle
   =
   \frac{\alpha^3}{3\pi}\ln\frac{\Lambda}{E_h}\>
   \left(
   4\pi Z\bigl\langle
      \delta(\mathbf{r})
   \bigr\rangle
   \right)
\end{equation}
\end{subequations}
which results in appearance of the logarithmic term, the cut-of parameter is later canceled out by the logarithmic contribution from the high energy part. Here $E_h$ is the Hartree energy.

The high energy part is obtained from the one-loop scattering amplitude for an electron in an external field \cite{Feyn49}
\begin{equation}\label{Feyn_amp}
\begin{array}{@{}l}
\displaystyle
M_1 =
   \frac{\alpha}{2\pi}
   \Bigl[
      2\left(\ln{\frac{m}{\lambda_{min}}}-1\right)
      \left(1-\frac{2\theta}{\tan{2\theta}}\right)
      +\theta\tan{\theta}
\\[4mm]\displaystyle\hspace{15mm}
      +\frac{4}{\tan{2\theta}}\int_0^\theta\alpha\tan{\alpha}\>d\alpha
   \Bigr]a_\nu\gamma^\nu
   +\frac{\alpha}{2\pi}
   \left[
      \frac{i}{2m}q_\mu a_\nu\Sigma^{\mu\nu}\frac{2\theta}{\sin{2\theta}}+r\,a_\nu\gamma^\nu
   \right],
\end{array}
\end{equation}
where
$\Sigma^{\mu\nu} = (\gamma^\mu\gamma^\nu\!-\!\gamma^\nu\gamma^\mu)/(2i)$,
$r=\ln(\lambda/m)\!+\!9/4\!-\!2\ln(m/{\lambda_{min}})$, and $q^2=4m^2\sin^2\theta$. Here amplitude is expressed in the natural relativistic units ($c=1$).

At small $q$, scattering amplitude may be expanded:
\[
\begin{array}{@{}l}\displaystyle
M_1 \approx
   \frac{\alpha}{\pi}
   \left[
      \left(-\frac{1}{8m^2}+\frac{1}{3m^2}\ln\frac{m}{\lambda_{min}}\right)
      a_\nu\gamma^\nu q^2
      +\frac{i}{4m}q_\mu a_\nu\Sigma^{\mu\nu}
   \right]
\\[3mm]\displaystyle\hspace{16mm}
   +\frac{\alpha}{\pi}
   \left[
      \left(-\frac{11}{240m^4}+\frac{1}{20m^4}\ln\frac{m}{\lambda_{min}}\right)
      a_\nu\gamma^\nu q^4
      +\frac{i}{24m^3}q_\mu a_\nu\Sigma^{\mu\nu}q^2
   \right]
\end{array}
\]

The leading order contribution for a \emph{static scalar field} with "renormalization"\/ to a new infrared regularization parameter $\lambda$, which is a cut-off of virtual quanta of momentum less than $\lambda$ ($\lambda=\alpha\Lambda$), is expressed
\begin{equation}\label{vertex1}
M_1^{(0)} =
   -\frac{\alpha}{3\pi}\frac{\mathbf{q}^2}{m^2}
   \left(\ln\frac{m}{2\lambda}+\frac{5}{6}-\frac{3}{8}\right) a_0
   +\frac{\alpha}{2\pi}\frac{1}{m^2}
      \left(-\frac{\mathbf{q}^2}{4}
      +\frac{i\boldsymbol{\sigma}[\mathbf{q\times p}]}{2}
   \right)a_0
\end{equation}
In order to get this expression, $\ln{m/\lambda_{min}}$ should be replaced by $[\ln{m/(2\lambda)}+5/6]$ \cite{Feyn49}. In the NRQED formalism this "renormalization" has been derived in \cite{Kin96}.

In the coordinate space and atomic units the energy displacement due to the respective interaction is expressed in the NRQED by
\begin{equation}\label{SE_high}
\Delta E_H =
   \frac{\alpha^3}{3\pi m^2}
   \left[
      \ln{\alpha^2}+\ln{\frac{\Lambda}{E_h}}+\ln{2}-\frac{5}{6}
   \right]
   \bigl\langle \Delta V \bigr\rangle
   -\frac{\alpha^3}{2\pi m^2}
   \left\langle
      \frac{\mathbf{r}\times\mathbf{p}}{r^3}\cdot\frac{\boldmath{\sigma}}{2}
   \right\rangle
\end{equation}

Summing up the low energy contribution (\ref{SE_fin}) and the high energy contribution (\ref{SE_high}), which comes from modification of the interaction of electron with a static field due to the one-loop self-energy diagram (Fig.~\ref{NRQED1}\,(a)), one gets
\begin{equation}\label{SE_contr}
\begin{array}{@{}l}\displaystyle
\Delta E_{SE} =
   \frac{\alpha^3}{3\pi m^2}
   \left[
      \ln{\alpha^2}+\ln{2}+\beta(n,l)-\frac{5}{6}
   \right]
   \bigl\langle \Delta V \bigr\rangle
   -\frac{\alpha^3}{2\pi m^2}
   \left\langle
      \frac{\mathbf{r}\times\mathbf{p}}{r^3}\cdot\frac{\boldmath{\sigma}}{2}
   \right\rangle
\end{array}
\end{equation}
where $\beta(n,l)$ is the so-called Bethe logarithm, it accumulates the contribution from (\ref{SE_fin}a) and will be formally defined later in Eq.~(\ref{bethe}). The cut-off parameter $\Lambda$ cancels out.

When both particles have finite masses one needs to include the retardation. The $\alpha(Z\alpha)^2(m/M)E_{nr}$ order low-energy contribution (Fig.~\ref{NRQED1}\,(b)) may be written (here $m_a$ and $m_b$ are masses of the two particles)
\begin{equation}\label{retard}
E_{L}^{ret}(a,b) = \frac{\alpha^3}{(4\pi)^2}\int_{|\mathbf{k}|<\Lambda}\frac{d\mathbf{k}}{k}
     \left(\delta^{ij}\!-\!\frac{k^ik^j}{k^2}\right)
     \left\langle
        \phi\left|\frac{p_a^i}{m_a}
           \left(
              \frac{1}{E_0\!-\!k\!-\!H_0}\!+\!{\frac{1}{k}}
           \right)
        \frac{p_b^j}{m_b}\right|\phi
     \right\rangle
\end{equation}
The last term in the inner round brackets, $-1/k$, corresponds to the retardation interaction as it appears in the Breit-Pauli approximation (and is of order $(Z\alpha)^2(m/M)E_{nr}$), and should be subtracted from the initial NRQED expression to avoid double counting. Consideration of the high-energy contribution ($k>\Lambda$), which comes from the same diagram, may be found in Ref.~\cite{Pac98}.

Summing up the contributions to the Bethe logarithm from Eqs.~(\ref{SE_fin}a) and (\ref{retard}) one finds that the dipole operator on the right and left-hand sides of the angle brackets may be recast as a nonrelativistic electric current density operator
\begin{equation}\label{current}
\mathbf{J}=\sum_i \frac{z_i}{m_i}\mathbf{P}_i,
\end{equation}
as it may be expected, since in the quantum electrodynamics a virtual photon interacts with a current density.

For a general many particle system the above speculations may be repeated directly in order to get the nonlogarithmic part of the low-energy contribution.

In summary, the Bethe logarithm may be defined as follows:

\noindent
Numerator:
\begin{subequations}\label{bethe}
\begin{equation}\label{numerator}
\mathcal{N}(L,v) =
   \int_0^{E_h}\,k\,dk
   \left\langle
      \mathbf{J}\left(\frac{1}{E_0\!-\!H\!-\!k}+\frac{1}{k}\right)\mathbf{J}
   \right\rangle
   +\int_{E_h}^\infty\,\frac{dk}{k}\,
   \left\langle
      \mathbf{J}\frac{(E_0\!-\!H)^2}{E_0\!-\!H\!-\!k}\mathbf{J}
   \right\rangle.
\end{equation}
Denominator:
\begin{equation}\label{denominator}
\mathcal{D}(L,v) =
   \Bigl\langle
      \mathbf{J}\left[H,\mathbf{J}\right]
   \Bigr\rangle =
   \frac{\bigl\langle
      \left[\mathbf{J}\left[H,\mathbf{J}\right]\right]
   \bigr\rangle}{2}\>.
\end{equation}
And the Bethe logarithm itself is a ratio of these two quantities
\begin{equation}\label{bethe_ratio}
\beta(L,v) = \frac{\mathcal{N}}{\mathcal{D}}\>.
\end{equation}
\end{subequations}

\subsection{First order perturbation wave function, $\psi_1(\cdot)$, and asymptotic expansion of $J(k)$ at $k\to\infty$.}

The key quantity for our numerical studies is
\begin{equation}\label{Jk}
J(k) =
   \left\langle
      \mathbf{J}\left(E_0\!-\!H\!-\!k\right)^{-1}\mathbf{J}
   \right\rangle.
\end{equation}
Knowing this function one immediately gets a value for the nonrelativistic Bethe logarithm using Eq.~(\ref{bethe}).

A general procedure to calculate $J(k)$ is to solve an equation
\begin{equation}\label{eq_psi1}
(E_0-H-k)\psi_1 = i\mathbf{J}\psi_0,
\end{equation}
for different values of $k$. Since we are interested in asymptotic behaviour of $J(k)$ for $k\to\infty$, it is assumed that $k$ is sufficiently large and as a first approximation one may take
\begin{equation}
\psi_1^{(0)} = -(i/k)\mathbf{J}\psi_0.
\end{equation}

Any approximate solution for $\psi_1$ may be (formally) iteratively improved:
\begin{equation}
\begin{array}{@{}l}\displaystyle
\psi_1^{(n)} =
   -\frac{i}{k}\mathbf{J}\psi_0+\frac{1}{k}(E_0\!-\!H)\psi_1^{(n-1)}
\end{array}
\end{equation}
and the next iteration would be
\begin{equation}\label{psi1^1}
\psi_1^{(1)} =
   -\frac{i}{k}\mathbf{J}\psi_0+\frac{1}{k^2}\left[H,i\mathbf{J}\right]\psi_0
\end{equation}
where
\begin{equation}\label{HJ}
[H,i\mathbf{J}] =
   \sum_{i>j} z_iz_j\left(\frac{z_j}{m_j}-\frac{z_i}{m_i}\right)\frac{\mathbf{r}_{ij}}{r_{ij}^3}\>,
\qquad
\mathbf{r}_{ij}=\mathbf{r}_j\!-\!\mathbf{r}_i.
\end{equation}

At small $r_{ij}$, $\psi_1$ should be smooth. In order to get a proper behaviour, one has to consider Eq.~(\ref{eq_psi1}) for $r_{ij}\to0$ and keep only important terms
\[
\left(\frac{1}{2m_{ij}}\Delta_{ij}-k\right)\psi_1(r_{ij},\cdot) = 0
\]
that gives homogeneous solutions of the type
\[
\sim\,\frac{\mathbf{r}_{ij}}{r_{ij}^3}\>e^{-\mu_{ij}r_{ij}}\,(1+\mu_{ij}r_{ij})
\]
with $\mu_{ij}=\sqrt{2m_{ij}k}$. These solutions, taken for different pairs of particles, may be added to $\psi_1^{(1)}$ to make the whole wave function smooth. So, we come to an approximation of $\psi_1$ for $k\!\to\!\infty$, which is of required quality for our aims and has the following form,
\begin{equation}
\psi_1^{(1)} =
   -\frac{i}{k}\mathbf{J}\psi_0(\cdot)
   +\frac{1}{k^2}\sum_{i>j}
      z_iz_j\left(\frac{z_j}{m_j}-\frac{z_i}{m_i}\right)
         \frac{\mathbf{r}_{ij}}{r_{ij}^3}
      \left[
         1-e^{-\mu_{ij}r_{ij}}\,(1+\mu_{ij}r_{ij})
      \right]\psi_0(\cdot)\,.
\end{equation}
As is seen from this equation, there is no singular term in the wave function corresponding to a pair of identical particles.

Integrand $J(k)$ may be evaluated using the variational formalism as a stationary solution of a functional on $\psi_1$
\[
J(k) =
   -2\left\langle \psi_0 |i\mathbf{J}| \psi_1 \right\rangle
   -\left\langle
      \psi_1 (E_0\!-\!H\!-\!k) \psi_1
   \right\rangle.
\]
To get asymptotic expansion we substitute $\psi_1^{(1)}$ into this functional. A derivation of the asymptotic expansion for the hydrogen ground state and comparison with known analytical result may be found in Appendix A.

At small $r_{ij}$ we get
\begin{subequations}\label{Jasy}
\begin{equation}
\begin{array}{@{}l}\displaystyle
J_{\rho_-} =
   -\left\langle
      \psi_1^{(1)} (E_0\!-\!H\!-\!k) \psi_1^{(1)}
   \right\rangle_{\rho_-}\!\!
 = -\frac{1}{k^3}\sum_{i>j}z_i^2z_j^2
   \left(\frac{z_i}{m_i}-\frac{z_j}{m_j}\right)^2\times
\\[4mm]\displaystyle\hspace{40mm}
   \Bigl[
      \sqrt{2m_{ij}k} +
      +z_iz_jm_{ij}\Bigl(\ln(m_{ij}k)\!-\!\ln{2}\!+\!1\!+\!2\gamma_E\!+\!2{\ln{\rho}}\Bigr)
   \Bigr]
   4\pi\left\langle\delta(\mathbf{r}_{ij})\right\rangle+\dots
\end{array}
\end{equation}
where $\rho_-$ means integration from $0$ to $\rho$. We assume that $\rho$ satisfies $1/\mu_{ij} \ll \rho \ll 1$.

For regular $r_{ij}$ we use $\psi_1^{(1)}$ in a form:
\[
\psi_1^{(1)} =
   -\frac{i}{k}\mathbf{J}\psi_0+\frac{1}{k^2}\left[H,i\mathbf{J}\right]\psi_0
\]
Then
\[
-2\left\langle \psi_0 |i\mathbf{J}| \psi_1^{(1)} \right\rangle =
   -\frac{2}{k}\left\langle\mathbf{J}^2\right\rangle
   -\frac{2}{k^2}\,\frac{
      \left\langle\>
         \left[i\mathbf{J}\!,\left[H,i\mathbf{J}\right]\right]
      \>\right\rangle}{2}
\]
and
\[
\begin{array}{@{}l}\displaystyle
   -\left\langle
      \psi_1^{(1)} (E_0\!-\!H\!-\!k) \psi_1^{(1)}
   \right\rangle_{\rho_+} =
   k\left\langle\psi_1^{(1)} \Big| \psi_1^{(1)}\right\rangle_{\rho_+}
   -\left\langle
      \psi_1^{(1)} (E_0\!-\!H) \psi_1^{(1)}
   \right\rangle_{\rho_+}
\\[4mm]\displaystyle\hspace{20mm}
 = \frac{1}{k}\left\langle\mathbf{J}^2\right\rangle
   +\frac{1}{k^2}\,\frac{
      \left\langle\>
         \left[i\mathbf{J}\!,\left[H,i\mathbf{J}\right]\right]
      \>\right\rangle}{2}
   -\frac{1}{k^3}
      \left[
         \bigl\langle\,
            \left[H,i\mathbf{J}\right]^2\,
         \bigr\rangle_{\rho_+}
         -\sum_{i>j}{\frac{z_i^2z_j^2\,m_{ij}}{\rho_{ij}}}\,
              4\pi\left\langle\delta(\mathbf{r}_{ij})\right\rangle
      \right]
\end{array}
\]
That results in
\begin{equation}
\widehat{J}_{\rho_+} =
   -\frac{1}{k}\left\langle\mathbf{J}^2\right\rangle
   -\frac{1}{k^2}\,\frac{
      \left\langle\>
         \left[i\mathbf{J}\!,\left[H,i\mathbf{J}\right]\right]
      \>\right\rangle}{2}
   -\frac{1}{k^3}
      \left[
         \bigl\langle\,
            \left[H,i\mathbf{J}\right]^2\,
         \bigr\rangle_{\rho_+}
         -\sum_{i>j}\frac{z_i^2z_j^2\,m_{ij}}{\rho_{ij}}\,
              4\pi\left\langle\delta(\mathbf{r}_{ij})\right\rangle
      \right]+\dots
\end{equation}
\end{subequations}

Now we have to introduce a finite functional, which should replace a divergent expectation value of the $1/r^4$ operator:
\begin{equation}\label{R}
\mathcal{R} =
   \lim_{\rho\to0}
   \left\{
   \left\langle
      \frac{1}{4\pi r^4}
   \right\rangle_{\!\!\rho}\!
   -\left[
      \frac{1}{\rho}\left\langle\delta(\mathbf{r})\right\rangle
      +\left(\ln{\rho}\!+\!\gamma_E\right)\left\langle\delta'(\mathbf{r})\right\rangle
   \right]
   \right\}
\end{equation}
where
\[
\left\langle\phi_1|\delta'(\mathbf{r})|\phi_2\right\rangle =
\left\langle\phi_1\left|
   \frac{\mathbf{r}}{r}\boldsymbol{\nabla}\delta(\mathbf{r})
\right|\phi_2\right\rangle =
   -\left\langle\partial_r\phi_1|\delta(\mathbf{r})|\phi_2\right\rangle
   -\left\langle\phi_1|\delta(\mathbf{r})|\partial_r\phi_2\right\rangle.
\]
Then summing up the Eqs.~(\ref{Jasy}a) and (\ref{Jasy}b) one gets
\begin{equation}\label{expansion}
\begin{array}{@{}l}\displaystyle
J(k) =
   -\frac{1}{k}\left\langle\mathbf{J}^2\right\rangle
   -\frac{1}{k^2}\,\frac{
      \left\langle\>
         \left[i\mathbf{J}\!,\left[H,i\mathbf{J}\right]\right]
      \>\right\rangle}{2}
   -\frac{1}{k^3}
      \sum_{\genfrac{}{}{0pt}{}{i>j,k>l}{(i,j)\ne (k,l)}}
         z_iz_jz_kz_l
         \left(\frac{z_i}{m_i}-\frac{z_j}{m_j}\right)
         \left(\frac{z_k}{m_k}-\frac{z_l}{m_l}\right)
         \left\langle\frac{\mathbf{r}_{ij}\mathbf{r}_{kl}}{r_{ij}^2r_{kl}^2}\right\rangle
\\[4mm]\displaystyle\hspace{12mm}
   -\frac{1}{k^3}
   \sum_{i>j}z_i^2z_j^2
   \left(\frac{z_i}{m_i}-\frac{z_j}{m_j}\right)^2
   \biggl\{
   4\pi\mathcal{R}_{ij}+
   \Bigl[
      \sqrt{2m_{ij}k}
      +z_iz_jm_{ij}\Bigl(\ln(m_{ij}k)\!-\!\ln{2}\!-\!1\Bigr)
   \Bigr]
   4\pi\left\langle\delta(\mathbf{r}_{ij})\right\rangle
   \biggr\}+\dots
\end{array}
\end{equation}

For mixed terms: $(\mathbf{r}_{ij}\mathbf{r}_{kl})/(r_{ij}r_{kl})^2$ in case of three-body calculation with the Hylleraas or exponential basis functions (see \cite{Kor00,Dra99}) a new type of singular integrals is required
\[
\Gamma_{-2,-2,n}(\alpha,\beta,\gamma) = \frac{1}{2}
   \int\int\>r_1^{-2}r_2^{-2}r_{12}^n\,e^{-\alpha r_1-\beta r_2 -\gamma r_{12}}\,
                 dr_1dr_2dr_{12}.
\]
A derivation of the explicit form for $\Gamma_{-2,-2,0}$ and stable recursions to get integrals for arbitrary $n$ are presented in Appendix B.

\subsection{Variational property}

If we consider a quantity
\begin{equation}
\mathcal{J}_\Lambda =
   \int_0^{\Lambda} k\,dk\>J(k) =
   \sum_n \bigl|\left\langle\psi_0|\mathbf{J}|\psi_n\right\rangle\bigr|^2
      \left[
         \Lambda-(E_0\!-\!E_n)\!\ln{\left|
                 \frac{E_0\!-\!E_n}{E_0\!-\!E_n\!-\!\Lambda}\right|}
      \right].
\end{equation}
we would find that for the ground state of a system this quantity possesses the variational property, since for the integrand for all $k$ the following inequality is fulfilled
\[
J_{\rm exact}(k) \ge J_{\rm numerical}(k).
\]
The same property remains satisfied for other states if integration is performed from some $k_0\sim 1$, which lies above the poles related to the states $E_n<E_0$. It is known from the practical calculations that the low $k$ contribution becomes numerically converged to a high accuracy at a moderate basis length of intermediate states, and thus with a good confidence the variational property, the higher the value of $\mathcal{J}_\Lambda$ the more accurate solution, is still remained in force. That allows us to perform optimization of the variational parameters of the basis set.

\section{Numerical results}

\subsection{Numerical scheme}

Here we consider the numerical scheme for the three-body Coulomb problem, which is then used in calculations of the Bethe logarithm for the helium and $\mbox{H}_2^+$ ground states. The wave functions both for the initial bound state and for the first order perturbation solution (or intermediate state), are taken in the form,
\begin{equation}\label{var}
\Psi_L(l_1,l_2) = \sum_{i=1}^{\infty}
    \Big\{
       U_i\,{\rm Re}\bigl[e^{-\alpha_i r_1-\beta_i r_2-\gamma_i r}\bigr]
      +W_i\,{\rm Im}\bigl[e^{-\alpha_i r_1-\beta_i r_2-\gamma_i r}\bigr]
\Big\}\mathcal{Y}^{l_1l_2}_{LM}(\mathbf{r}_1,\mathbf{r}_2),
\end{equation}
where $\mathcal{Y}^{l_1l_2}_{LM}(\mathbf{r}_1,\mathbf{r}_2)$ are the solid bipolar harmonics as defined in \cite{Varshalovich}, $L$ is a total orbital angular momentum of a state. Complex parameters $\alpha_i$, $\beta_i$ and $\gamma_i$ are generated in a quasi-random manner
\cite{Kor00}:
\begin{equation}
\begin{array}{l}\displaystyle
\alpha_i =
   \left[\left\lfloor
            \frac{1}{2}i(i+1)\sqrt{p_\alpha}
         \right\rfloor(A_2-A_1)+A_1\right]
 +i\left[\left\lfloor\frac{1}{2}i(i+1)\sqrt{q_\alpha}
         \right\rfloor(A'_2-A'_1)+A'_1\right],
\end{array}
\end{equation}
$\lfloor x\rfloor$ designates the fractional part of $x$, $p_\alpha$ and $q_\alpha$ are some prime numbers, $[A_1,A_2]$ and $[A'_1,A'_2]$ are real variational intervals which need to be optimized. Parameters $\beta_i$ and $\gamma_i$ are obtained in a similar way.

Basis set for intermediate states is constructed as follows:
\begin{enumerate}
\item
First we use a regular basis set, which is taken similarly to the initial state with regular values of parameters $(\alpha,\beta,\gamma)$ in exponentials.
\item Then we build a special basis set with exponentially growing parameters for a particular $r_{ij}$
\begin{equation}\label{basis_2}
\left\{
\begin{array}{@{}ll}\displaystyle
A_1^{(0)} = A_1, & A_2^{(0)} = A_2
\\[1mm]\displaystyle
A_1^{(n)} = \tau^n A_1, \qquad & A_2^{(n)} = \tau^n A_2
\end{array}\right.
\end{equation}
where $\tau=A_2/A_1$.

Typically $[A_1,A_2] = [2.5,4.5]$, and $n_{\rm max} = 5\!-\!7$, that corresponds to the photon energy interval $k\in[0,10^4]$.
\item For other pairs of $(i,j)$ we take the similar basis sets as in 2. It is worthy to note that for identical particles this step should be omitted as is discussed in previous section.
\end{enumerate}

After the complete set of basis functions is constructed, we diagonalize matrix of the Hamiltonian $H_I$ for intermediate states to get a set of (pseudo)state energies, $E_m$, and then to calculate $\left\langle 0 |i\mathbf{J}| m \right\rangle$. These two sets of data are enough to restore $J(k)$:
\begin{equation}
J(k) = -\sum_m
     \frac{\left\langle 0| i\mathbf{J} |m\right\rangle^2}{E_0\!-\!E_m\!-\!k}\>,
\end{equation}
and to integrate the low energy part of the numerator $\mathcal{N}(L,v)$
\begin{equation}
\int_0^{E_h}\,k\,dk
   \left\langle
      \mathbf{J}\left(\frac{1}{E_0\!-\!H\!-\!k}+\frac{1}{k}\right)\mathbf{J}
   \right\rangle
   +\int_{E_h}^\Lambda\,\frac{dk}{k}\,
   \left\langle
      \mathbf{J}\frac{(E_0\!-\!H)^2}{E_0\!-\!H\!-\!k}\mathbf{J}
   \right\rangle.
\end{equation}

From thus obtained $J(k)$ we extrapolate coefficients of asymptotic expansion
\begin{equation}\label{asy_num}
f_{\rm fit}(k)=\sum_{m=1}^M \frac{C_{1m}\sqrt{k}\!+\!C_{2m}\ln{k}\!+\!C_{3m}}{k^{m+3}}
\end{equation}
which is taken in the same form as in analytic expression for the hydrogen atom (see Appendix A, Eq.~(\ref{H_asy})). The similar asymptotic expansion has been used in \cite{Yer09}. The leading order terms of $J(k)$ are obtained from Eq. (\ref{expansion}). That allows to get the high energy part of the numerator
\[
\int_\Lambda^\infty\,\frac{dk}{k}\,
   \left\langle
      \mathbf{J}\frac{(E_0\!-\!H)^2}{E_0\!-\!H\!-\!k}\mathbf{J}
   \right\rangle\>.
\]

\begin{table}[t]
\begin{tabular}{r@{\hspace{5mm}}l@{\hspace{10mm}}l@{\hspace{5mm}}r}
\hline\hline
\multicolumn{2}{c}{this work} & \multicolumn{2}{c}{\protect\cite{Dra99}} \\
\hline
 $N$ & \hspace{12mm}$\beta$ & \hspace{7mm}$\beta$ & $N$ \\
\hline
 40 & $2.98412855576549_0$      & $2.9841284_9$       & 45  \\
 60 & $2.984128555765497_3$     & $2.984128555_1$     & 66  \\
 80 & $2.98412855576549760_7$   & $2.98412855575_9$   & 91  \\
100 & $2.9841285557654976107_3$ & $2.9841285557654_4$ & 120 \\
\hline
\multicolumn{4}{c}{exact\hspace{5mm} 2.98412855576549761075977709002} \\
\hline\hline
\end{tabular}
\caption{Convergence of the Bethe logarithm for the ground state of hydrogen and comparison with results of Drake and Goldman \cite{Dra99}.}\label{H_bethe}
\end{table}
\begin{figure}[t]
\includegraphics[width=0.5\textwidth]{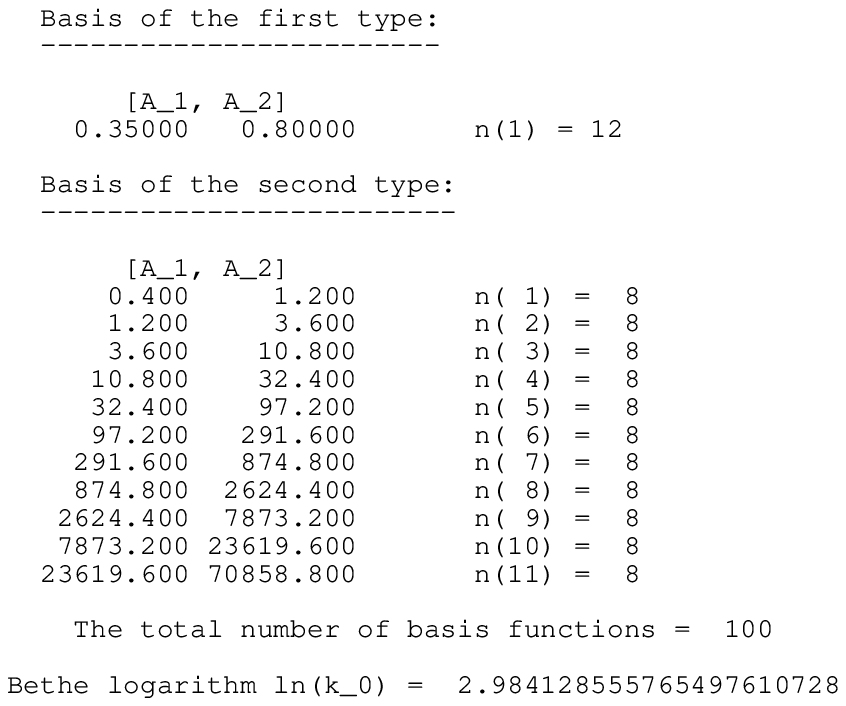}
\caption{Parameters of the basis set and numerical value of the Bethe logarithm for the ground state of hydrogen for $N=100$.}\label{H_bethe_lst}
\end{figure}

\subsection{Results}

As a first example demonstrating capabilities of our method we take the ground state of a hydrogen atom. Results of numerical calculation are summarized in Table \ref{H_bethe}. A basis set used for these calculations is a sum of exponentials with real parameters generated in a quasi-random way, the initial wave function is taken exact. On Fig.~\ref{H_bethe_lst} an excerption of the listing of output with the parameters of the basis set for the case of $N=100\,$ functions is shown. "Exact" in Table \ref{H_bethe} means the value obtain by Huff's method \cite{Huff}, which is a series expansion. It is interesting to note that if the Huff series are taken with 100 terms only the result would be of the same accuracy as in our approach for $N=100$. As is seen from this Table the new method demonstrates better accuracy than in \cite{Dra99}, especially for small $N$, and good convergence rate. Worthy to note that variational parameters were roughly optimized for $N=80$, and kept the same for all other cases.

The main results of our studies are presented in Tables  \ref{He_bethe} and \ref{H2_bethe}. The first is the Bethe logarithm calculations for the ground state of helium and the next table is for the ground state of $\mbox{H}_2^+$ molecular ion. Convergence is analyzed in two dimensions, with respect to the number of basis functions of the initial state and similarly for the basis size of the intermediate state. The major conclusion is that for ultimate precision it is important to check how the studied quantity depends on accuracy of the initial state. Indeed, it is reasonable to expect that the fractional error in the Bethe logarithm evaluation would be no better than the square root of the fractional uncertainty in the variational energy of the initial state, like in behaviour of expectation values of, say, $\delta$-function or $\mathbf{p}^4$ operators.

\begin{table}
\begin{tabular}{c@{\hspace{7mm}}l@{\hspace{7mm}}l@{\hspace{7mm}}l@{\hspace{7mm}}l}
\hline\hline
$N_b$ $\backslash$ $N_a$ & ~~~~~3000 & ~~~~~3500 & ~~~~~4000 & $~~~~~~\infty$ \\
\hline
 4000 & 4.37016022311 & 4.37016022301 & 4.370160223021 \\
 5000 & 4.37016022314 & 4.37016022303 & 4.370160223044 \\
 6000 &               & 4.37016022304 & 4.370160223058 \\
\hline
$\infty$&            & & & 4.37016022306(2) \\
\hline\hline
\end{tabular}
\caption{Test of convergence of the Bethe logarithm value for the ground state of a helium atom. $N_a$ is a basis length for the initial state, $N_b$ is a basis length for an intermediate state.}\label{He_bethe}
\end{table}

\begin{table}
\begin{tabular}{c@{\hspace{7mm}}l@{\hspace{7mm}}l@{\hspace{7mm}}l@{\hspace{7mm}}l}
\hline\hline
$N_b$ $\backslash$ $N_a$ & ~~~~~3000 & ~~~~~4000 & ~~~~~5000 & ~~~~~~$\infty$ \\
\hline
 7000 & 3.0122303407 & 3.0122303334 & \\
 8000 & 3.0122303431 & 3.0122303357 & 3.0122303341 \\
 9000 & 3.0122303442 & 3.0122303367 & 3.0122303349 \\
\hline
 $\infty$ &         &              &              & 3.012230335(1) \\
\hline\hline
\end{tabular}
\caption{Test of convergence of the Bethe logarithm value for the ground state of a hydrogen molecular ion $\mbox{H}_2^+$. $N_a$ is a basis length for the initial state, $N_b$ is a basis length for an intermediate state.}\label{H2_bethe}
\end{table}

Comparing our result for the ground state of helium with \cite{Kor04} based on Goldman-Drake method (see Table \ref{comparison}) we see a discrepancy which requires to be somehow explained. This puzzle was a serious challenge for us, since the study of convergence (see Table II of \cite{Kor04}) unambiguously showed that the stability is achieved. So, we came back to our old calculations, which were based on the Goldman-Drake approach, and it was found that some essential part in the intermediate state wave function had been missed. If we add a new set of "regular" basis functions with exponentials having parameters $\alpha$ and $\beta$ (in front of $r_1$ and $r_2$, respectively) up to 40(!) then the numerical result become $\beta=4.370\,160\,222\,67$ for the case of initial $N=1400$ state used in \cite{Kor04}. That showed a good coincidence with our "velocity gauge" result for the same initial state, as it might be expected!

For convenience of comparison with other calculations we present here explicitly the asymptotic expansion of $J(k)$ ($k\to\infty$) for the helium ground state:
\begin{equation}
\begin{array}{@{}l}\displaystyle
J(k)+\frac{1}{k}\left\langle\mathbf{J}^2\right\rangle =
   \frac{4\pi
   \left[
      \left\langle\delta(\mathbf{r}_1)\right\rangle
      \!+\!\left\langle\delta(\mathbf{r}_2)\right\rangle
   \right]}{k^2}
   \biggl\{
      \frac{Z}{2}
      -\frac{Z^2\sqrt{2k}-Z^3\ln{k}+20.00249948}{k}
\\[3mm]\displaystyle\hspace{65mm}
      +\frac{80.3063\sqrt{k}-70.989\ln{k}+136.5}{k^2}+\dots
   \biggr\}.
\end{array}
\end{equation}
Numerical coefficient in the first line has been calculated using Eq.~(\ref{expansion}). Coefficients shown in a second line were obtained by a linear least squares fit using the SVD algorithm \cite{LH}. The fitting interval was $k\in[20,6000]$, number of data points was 100, which were taken equidistant on the logarithmic scale of $k$, number of terms in the asymptotic expansion is 14-18. The results of fitting procedure are not sensitive to a number of data point.

\begin{table}[t]
\begin{tabular}{l@{\hspace{6mm}}l}
\hline\hline
 & $~~~~~\beta$ \\
 \hline
Schwartz \cite{Schwartz}           & 4.370(4)      \\
Korobov and Korobov \cite{Kor99}   & 4.370\,158(1) \\
Baker {\em et al.}., \cite{Baker}  & 4.370\,159(2) \\
Drake and Goldman \cite{Dra99}     & 4.370\,160\,218(3)   \\
Korobov \cite{Kor04}               & 4.370\,160\,2220(1)  \\
this work                          & 4.370\,160\,22306(2) \\
\hline\hline
\end{tabular}
\caption{Calculations of the Bethe logarithm for the helium ground state.}\label{comparison}
\end{table}

Similarly, the asymptotic expansion of $J(k)$ for the ground state of a molecular ion $\mbox{H}_2^+$ is
\begin{equation}
\begin{array}{@{}l}\displaystyle
J(k)+\frac{1}{k}\left\langle\mathbf{J}^2\right\rangle =
   \frac{2\pi
   \left[
      Z_1\left\langle\delta(\mathbf{r}_1)\right\rangle
      \!+\!Z_2\left\langle\delta(\mathbf{r}_2)\right\rangle
   \right]}{k^2}
   -\left(\frac{1}{m_e}\!+\!{Z_1}{M_1}\right)^2
   \frac{\left[Z_1^2\sqrt{2\mu_1k}-Z_1^3\mu_1\ln{\mu_1k}\right]
                            4\pi\left\langle\delta(\mathbf{r}_1)\right\rangle
         }{k^3}
\\[3mm]\displaystyle\hspace{67mm}
   -\left(\frac{1}{m_e}\!+\!{Z_2}{M_2}\right)^2
   \frac{\left[Z_2^2\sqrt{2\mu_2k}-Z_2^3\mu_2\ln{\mu_2k}\right]
                            4\pi\left\langle\delta(\mathbf{r}_2)\right\rangle
         }{k^3}
\\[3.5mm]\displaystyle\hspace{27mm}
   -2\pi
   \left[
      Z_1\left\langle\delta(\mathbf{r}_1)\right\rangle
      \!+\!Z_2\left\langle\delta(\mathbf{r}_2)\right\rangle
   \right]
   \biggl\{
      \frac{2.24754280}{k^3}
      +\frac{10.052\sqrt{k}-4.4\ln{k}+2.2}{k^4}+\dots
   \biggr\}.
\end{array}
\end{equation}
where $\mu_i=m_eM_i/(m_e\!+\!M_i)$ is the reduced mass of a respective proton of mass $M_i$. Since $\bigl\langle\delta(\mathbf{r}_{12})\bigr\rangle<10^{-10}$, the term which is related to a nucleus-nucleus part (say, for HD$^+$ case) may be neglected.

One additional remark on $\mbox{H}_2^+$ molecular ion is needed. Analyzing our new results for the Bethe logarithm we found that they have a systematic shift compared to the ones of \cite{H2_BL}. That is because the formula for asymptotic expansion used in previous calculations (see Ref.~\cite{HD_BL}, Eq.~(12)) does not incorporate reduced masses into the leading terms of the expansion. This shift is almost state independent and (both for $\mathrm{H}_2^+$ \cite{H2_BL} and $\mathrm{HD}^+$ \cite{HD_BL}) is of about 15 in the last two digits indicated in the Tables of \cite{H2_BL,HD_BL} and should be subtracted. A new systematic calculation of the Bethe logarithm for the hydrogen molecular ions, which should provide better accuracy of about 8-9 significant digits, is in progress now.

\section{Acknowledgements}

The author want to express his thanks to K.~Pachucki, V.A.~Yerokhin, L.~Hilico, and J.-Ph.~Karr for helpful discussions. Comments on the manuscript by Z.-X.~Zhong are greatly appreciated. The work was supported in part by the Russian Foundation for Basic Research, Grant No.~12-02-00417-a. I also want to acknowledge support of Chinese Academy of Science (CAS) during my stay at WIPM, Wuhan Branch of CAS.

\appendix

\section{Asymptotic expansion of $J(k)$: Hydrogen ground state.}

Here the Hamiltonian and the charge current density operator are expressed
\[
H = -\frac{\boldsymbol{\nabla}^2}{2}-\frac{Z}{r}\,,
\qquad
i\mathbf{J} = -\boldsymbol{\nabla}\,.
\]
The following commutations are helpful for our derivation
\[
[H,\boldsymbol{\nabla}] =
   -Z\frac{\mathbf{r}}{r^3}\>,
\qquad
\left[\boldsymbol{\nabla},\left[H,\boldsymbol{\nabla}\right]\right] =
   -4\pi Z \left\langle\delta(\mathbf{r})\right\rangle\>.
\]
We substitute the wave function
\[
\begin{array}{@{}l}\displaystyle
\psi_1^{(1)}(\mathbf{r}) =
   \frac{1}{k}\boldsymbol{\nabla}\psi_0(r)
   -\frac{1}{k^2}
      \left[H,\boldsymbol{\nabla}\right]
      \left[
         1-e^{-\mu r}\,(1+\mu r)
      \right]\psi_0(r)\,.
\\[3mm]\displaystyle\hspace{12mm}
 = -\frac{Z}{k}\,\psi_0(r)
   +\frac{Z^2\mathbf{r}}{k^2r^3}
      \left[
         1-e^{-\mu r}\,(1+\mu r)
      \right]\psi_0(r)\,.
\end{array}
\]
where $\psi_0=2Z^{3/2}e^{-Zr}$ is the ground state wave function and $\mu=\sqrt{2k}$, into the variational functional
\[
J(k) =
   -2\left\langle \psi_0 |\boldsymbol{\nabla}| \psi_1 \right\rangle
   -\left\langle
      \psi_1 (E_0\!-\!H\!-\!k) \psi_1
   \right\rangle
\]

\begin{itemize}
\item For $r\!<\!\rho$ ($\rho\!\to\!0$, and $\mu\rho\!\gg\!0$) one gets:
\[
J_{\rho_-} =
   4\pi\left\langle\delta(\mathbf{r})\right\rangle\>Z^{-3}
   \left[
      -\frac{Z^5\sqrt{2k}}{k^3}
      +\frac{Z^6(\ln{k}-\ln{2}+1)}{k^3}
      +\frac{2Z^6(\gamma_E+\ln{\rho})}{k^3}
   \right]
\]

\item For $r>\rho$:
\[
\begin{array}{@{}l}\displaystyle
J_{\rho_+} =
   \frac{1}{k}\left\langle\boldsymbol{\nabla}^2\right\rangle
   -\frac{1}{k^2}\,\frac{
      \left\langle\>
         \left[\boldsymbol{\nabla}\!,\left[H,\boldsymbol{\nabla}\right]\right]
      \>\right\rangle}{2}
   -\frac{1}{k^3}
      \left[
         \bigl\langle\,
            \left[H,\boldsymbol{\nabla}\right]^2\,
         \bigr\rangle_{\rho_+}
         -\frac{Z^2}{\rho}\,4\pi\left\langle\delta(\mathbf{r})\right\rangle
      \right]
\\[3mm]\hspace{10mm}\displaystyle
 = -\frac{Z^2}{k}+\frac{2Z^4}{k^2}
   -\frac{8Z^6\left[\gamma_E+\ln(2Z\rho)\right]}{k^3}+\dots
\end{array}
\]
\end{itemize}
Summing up, we obtain the leading terms of the asymptotic expansion:
\begin{equation}
\begin{array}{@{}l}\displaystyle
J(k) =
   \frac{1}{k}\left\langle\boldsymbol{\nabla}^2\right\rangle
   -\frac{1}{k^2}\,\frac{
      \left\langle\>
         \left[\boldsymbol{\nabla}\!,\left[H,\boldsymbol{\nabla}\right]\right]
      \>\right\rangle}{2}
   -\frac{1}{k^3}
      \left[
         \bigl\langle\,
            \left[H,\boldsymbol{\nabla}\right]^2\,
         \bigr\rangle_{\rho_+}
         -\frac{Z^2}{\rho}\,4\pi\left\langle\delta(\mathbf{r})\right\rangle
      \right]
\\[2mm]\hspace{20mm}\displaystyle
   +\left[
      -\frac{Z^5\sqrt{2k}}{k^3}
      +\frac{Z^6(\ln{k}-\ln{2}+1)}{k^3}
      +\frac{2Z^6(\gamma_E+\ln{\rho})}{k^3}
   \right]
   Z^{-3}4\pi\left\langle\delta(\mathbf{r})\right\rangle\>
\\[3mm]\hspace{9mm}\displaystyle
 = -\frac{Z^2}{k}+\frac{2Z^4}{k^2}
   -\frac{4Z^5\sqrt{2k}}{k^3}+\frac{4Z^6(\ln{k}-\ln{Z^2})}{k^3}
   -\frac{4Z^6(3\ln{2}-1)}{k^3}+\dots
\end{array}
\end{equation}
which is exact up to free term in the $1/k^3$ order as it may be checked from comparison with analytical expression known for this case \cite{Pac93} ($Z\!=\!1$):
\begin{equation}\label{H_asy}
\begin{array}{@{}l}\displaystyle
J(k) =
   -384\>\frac{\tau^5}{(1+\tau)^8(2-\tau)} \; {}_2F_1(4,2-\tau,3-\tau;\xi)
\\[3mm]\displaystyle\hspace{8mm}
 = -\frac{1}{k}+\frac{2}{k^2}-\frac{4\sqrt{2k}}{k^3}
   +\frac{4\ln{k}\!-\!12\ln{2}\!+\!4}{k^3}
   +\frac{(21\!+\!2\pi^2)\sqrt{2k}}{3k^4}
   -\frac{8\ln{k}\!-\!24\ln{2}\!+\!14\!+\!2\psi''(2)}{k^4}+\dots
\end{array}
\end{equation}
where $\tau = Z/\sqrt{-2(E_0-k)}$ and $\xi = [(1-\tau)/(1+\tau)]^2$.

\section{$\Gamma_{-2,-2,n}(\alpha,\beta,\gamma)$}

In this appendix we show how to evaluate the integral
\begin{equation}\label{integral_n}
\Gamma_{-2,-2,n}(\alpha,\beta,\gamma) = \frac{1}{2}
   \int\int\>r_1^{-2}r_2^{-2}r_{12}^n\,e^{-\alpha r_1-\beta r_2 -\gamma r_{12}}\,
                 dr_1dr_2dr_{12},
\end{equation}
by means of
\begin{equation}\label{diff-rel}
\Gamma_{-2,-2,n}(\alpha,\beta,\gamma) =
   \left(-\frac{\partial}{\partial \gamma}\right)^n
      \Gamma_{-2,-2,0}(\alpha,\beta,\gamma).
\end{equation}

To do this we have to obtain an analytical expression for $\Gamma_{-2,-2,0}(\alpha,\beta,\gamma)$. Let us consider
\[
\begin{array}{@{}l}\displaystyle
\Gamma_{-2,-2,0}(\alpha,\beta,\gamma) =
   \frac{1}{2}\int_{\epsilon_1}^{\infty} dr_1 \frac{e^{-\alpha r_1}}{r_1^2}
      \Biggl[
      \int_{\epsilon_2}^{r_1}\>dr_2\frac{e^{-\beta r_2}}{r_2^2}
         \int_{r_1-r_2}^{r_1+r_2}dr_{12}e^{-\gamma r_{12}}
\\[3mm]\displaystyle\hspace{80mm}
      +\int_{r_1}^{\infty}\>dr_2\frac{e^{-\beta r_2}}{r_2^2}
         \int_{r_2-r_1}^{r_1+r_2}dr_{12}e^{-\gamma r_{12}}
      \Biggr]
\\[5mm]\displaystyle\hspace{10mm}
 = \frac{1}{2\gamma}\int_{\epsilon_1}^{\infty} dr_1
                \frac{e^{-(\alpha+\gamma) r_1}}{r_1^2}
      \int_{\epsilon_2}^{\infty}\>\frac{dr_2}{r_2^2}
         \left(e^{-(\beta-\gamma)r_2}-e^{-(\beta+\gamma)r_2}\right)
\\[3mm]\displaystyle\hspace{30mm}
   -\frac{1}{2\gamma}\int_{\epsilon_1}^{\infty} dr_1
       \left[
          \frac{e^{-(\alpha+\beta)r_1}}{r_1^3}
          -(\beta-\gamma)\frac{e^{-(\alpha+\gamma)r_1}
               E_1\bigl((\beta\!-\!\gamma)r_1\bigr)}{r_1^2}
       \right]
\\[3mm]\displaystyle\hspace{30mm}
   +\frac{1}{2\gamma}\int_{\epsilon_1}^{\infty} dr_1
       \left[
          \frac{e^{-(\alpha+\beta)r_1}}{r_1^3}
          -(\beta+\gamma)\frac{e^{-(\alpha-\gamma)r_1}
               E_1\bigl((\beta\!+\!\gamma)r_1\bigr)}{r_1^2}
       \right]
\\[4mm]\displaystyle\hspace{10mm}
 = \frac{1}{2\gamma}\,L_2(\alpha\!+\!\gamma,\epsilon_1)
         \Bigl[L_2(\beta\!-\!\gamma,\epsilon_2)-L_2(\beta\!+\!\gamma,\epsilon_2)\Bigr]
\\[3mm]\displaystyle\hspace{40mm}
   +\frac{\beta\!-\!\gamma}{2\gamma}
      I_2(\alpha+\gamma,\beta\!-\!\gamma,\epsilon_1)
   -\frac{\beta\!+\!\gamma}{2\gamma}
      I_2(\alpha\!-\!\gamma,\beta\!+\!\gamma,\epsilon_1)\>.
\end{array}
\]
Here we use the notation from \cite{Har04}:
\[
\begin{array}{@{}l}\displaystyle
L_p(x,\epsilon) = \int_{\epsilon}^{\infty} \frac{e^{-x t}}{t^p}\>dt =
   x^{p-1}\Gamma(1\!-\!p, x\epsilon),
\\[2mm]\displaystyle\hspace{15mm}
L_1(x,\epsilon) = \psi(1)-\ln(x\epsilon),
\qquad
L_2(x,\epsilon) =
   \frac{1}{\epsilon}-x\left[\psi(2)-\ln(x\epsilon)\right]-\frac{x^2\epsilon}{2},
\\[4mm]\displaystyle
I_p(x,y,\epsilon) = \int_{\epsilon}^{\infty} \frac{e^{-x t}}{t^p}\,E_1(yt)\>dt,
\\[4mm]\displaystyle\hspace{15mm}
I_p(x,y,\epsilon) =
   \frac{e^{-x\epsilon}E_1(y\epsilon)}{(p\!-\!1)\epsilon^{p-1}}
   -\frac{x}{p-1}I_{p-1}(x,y)-\frac{1}{p-1}L_p(x+y,\epsilon),
\\[4mm]\displaystyle\hspace{15mm}
I_1(x,y,\epsilon) =
   \frac{1}{2}\left(\ln{\epsilon}\!+\!\ln{y}\!+\!\gamma_E\right)^2
   -\frac{\pi^2}{12}-\frac{1}{2}\ln^2{\frac{y}{x}}
   -\mathrm{dilog}\left(1+\frac{y}{x}\right).
\end{array}
\]
In expressions above $\psi(n)$ is a digamma function: $\psi(1)=-\gamma_E$, $~\psi(n\!+\!1)=\psi(n)+1/n\,$.

The integral is divergent and in order to get some meaningful finite expression one has to introduce some counterterm similar to what was done for the functional $\mathcal{R}$ in Eq.~(\ref{R}). Still for our aims it is not needed, since we will be using $\Gamma_{-2,-2,0}$ only for evaluation of finite integrals and, thus, any form of  $\Gamma_{-2,-2,0}$, which is self-consistent with other $\Gamma_{lmn}(\alpha,\beta,\gamma)$ via differentiation relations like (\ref{diff-rel}), would be sufficient. So we may choose it in the following form:
\begin{equation}
\begin{array}{@{}l}\displaystyle
\Gamma_{-2,-2,0}(\alpha,\beta,\gamma) =
   \left(3-2\gamma_E-\frac{\pi^2}{12}\right)\gamma
   -(\alpha+\beta)
\\[3mm]\hspace{20mm}\displaystyle
   -\frac{\alpha\!+\!\gamma}{2}\Bigl(2-\gamma_E-\ln(\alpha\!+\!\gamma)\Bigr)^2
   -\frac{\beta\!+\!\gamma}{2}\Bigl(2-\gamma_E-\ln(\beta\!+\!\gamma)\Bigr)^2
\\[3mm]\hspace{20mm}\displaystyle
   +(\alpha+\beta)\ln(\alpha+\beta)
   -(\alpha+\gamma)\ln(\alpha+\gamma)-(\beta+\gamma)\ln(\beta+\gamma)
\\[3mm]\hspace{20mm}\displaystyle
   +\frac{(\alpha+\gamma)(\beta+\gamma)}{4\gamma}
                     \ln^2\left(\frac{\alpha+\gamma}{\beta+\gamma}\right)
   +\frac{\alpha\beta\pi^2}{12\gamma}
\\[3mm]\hspace{20mm}\displaystyle
   +\frac{(\alpha-\gamma)(\beta+\gamma)}{2\gamma}\>
                    \mbox{dilog}\left(\frac{\alpha+\beta}{\beta+\gamma}\right)
   +\frac{(\beta-\gamma)(\alpha+\gamma)}{2\gamma}\>
                    \mbox{dilog}\left(\frac{\alpha+\beta}{\alpha+\gamma}\right)\>.
\end{array}
\end{equation}
Evaluation of $\Gamma_{-2,-2,n}$ is straightforward (see \cite{JPB02} for details) except for two terms, which require additional remarks. We introduce two recursions:
\begin{equation}
A_n=\left(-\frac{\partial}{\partial\gamma}\right)^n
   \left[
      \frac{1}{\gamma}\>\mbox{dilog}\left(\frac{\alpha+\beta}{\alpha+\gamma}\right)
   \right] =
   \frac{1}{\gamma}\left[
      n A_{n-1}+B_n
   \right],
\qquad B_n = \left(-\frac{\partial}{\partial\gamma}\right)^n
   \left[
      \mbox{dilog}\left(\frac{\alpha+\beta}{\alpha+\gamma}\right)
   \right],
\end{equation}
and
\begin{equation}
E_n = \left(-\frac{\partial}{\partial\gamma}\right)^n
   \left[
      \frac{1}{\gamma}\>\ln^2\left(\frac{\alpha+\gamma}{\beta+\gamma}\right)
   \right] =
   \frac{1}{\gamma}\left[
      n E_{n-1}+F_n
   \right],
\qquad
F_n = \left(-\frac{\partial}{\partial\gamma}\right)^n
   \left[
      \ln^2\left(\frac{\alpha+\gamma}{\beta+\gamma}\right)
   \right].
\end{equation}
At first glance these recursions are not stable, when $\gamma$ is small. However, more careful analysis shows that, say, $A_{n-1}$ and $B_n$ (as well as $E_{n-1}$ and $F_n$) are of the same sign, and no subtraction, which leads to loss of numerical accuracy, occurs.

\end{document}